\input harvmac.tex
 \input epsf.tex
 \input amssym

\def\figin{\epsfcheck\figin}\def\figins{\epsfcheck\figins}
\def\epsfcheck{\ifx\epsfbox\UnDeFiNeD
\message{(NO epsf.tex, FIGURES WILL BE IGNORED)}
\gdef\figin##1{\vskip2in}\gdef\figins##1{\hskip.5in}%
\else\message{(FIGURES WILL BE INCLUDED)}%
\gdef\figin##1{##1}\gdef\figins##1{##1}\fi}
\def\DefWarn#1{}
\def\figinsert{\goodbreak\midinsert}
\def\ifig#1#2#3{\DefWarn#1\xdef#1{fig.~\the\figno}
\writedef{#1\leftbracket fig.\noexpand~\the\figno} %
\figinsert\figin{\centerline{#3}}\medskip\centerline{\vbox{\baselineskip12pt
\advance\hsize by -1truein\noindent\footnotefont{\bf
Fig.~\the\figno:} #2}}
\bigskip\endinsert\global\advance\figno by1}
\def\writetoci{\immediate\openout\tfile=toc.tmp
   \def\writetoca##1{{\edef\next{\write\tfile{\noindent ##1 
   \string\leaderfill {\noexpand\number\pageno} \par}}\next}}}
\def\writetocii{\immediate\openout\tfile=toc.tex
   \def\writetoca##1{{\edef\next{\write\tfile{\noindent ##1 
   \string\leaderfill {\noexpand\number\pageno} \par}}\next}}}

\def\unit{\relax{\rm 1\kern-.26em I}}
\def\nada{\relax{\rm 0\kern-.30em l}}

\def \pa {\partial}

\def \eps {\epsilon}
\def\bz{{\bar z}}
\def \e {\epsilon}
\def\im{{\rm Im}}
\def\p{\partial}

\noblackbox
\def\IL{\relax{\rm I\kern-.18em L}}
\def\IH{\relax{\rm I\kern-.18em H}}
\def\IR{\relax{\rm I\kern-.18em R}}
\def\IC{\relax\hbox{$\inbar\kern-.3em{\rm C}$}}
\def\IZ{\relax\ifmmode\mathchoice
{\hbox{\cmss Z\kern-.4em Z}}{\hbox{\cmss Z\kern-.4em Z}} {\lower.9pt\hbox{\cmsss Z\kern-.4em Z}}
{\lower1.2pt\hbox{\cmsss Z\kern-.4em Z}}\else{\cmss Z\kern-.4em Z}\fi}

\def\ci {{\cal I }}

\def\CO {{\cal O}}

\def\CC {{\cal C}}

\def\CA{{\cal A}}


\def\CO {{\cal O}}

\def\zb {\bar{z}}

\font\manual=manfnt \def\dbend{\lower3.5pt\hbox{\manual\char127}}

\def\ip{${\cal I}^+$}

\def\re{{\rm Re}}

\lref\bt{
  G.~Barnich and C.~Troessaert,
  ``BMS charge algebra,''
  JHEP {\bf 1112}, 105 (2011)
  [arXiv:1106.0213 [hep-th]].
    }
\lref\StromingerJFA{
  A.~Strominger,
  ``On BMS Invariance of Gravitational Scattering,''
JHEP {\bf 1407}, 152 (2014).
[arXiv:1312.2229 [hep-th]].
}

\lref\BieriADA{
  L.~Bieri and D.~Garfinkle,
  ``A perturbative and gauge invariant treatment of gravitational wave memory,''
Phys.\ Rev.\ D {\bf 89}, 084039 (2014).
[arXiv:1312.6871 [gr-qc]].
}

\lref\TolishBKA{
  A.~Tolish and R.~M.~Wald,
  ``Retarded Fields of Null Particles and the Memory Effect,''
[arXiv:1401.5831 [gr-qc]].
}

\lref\TolishODA{
  A.~Tolish, L.~Bieri, D.~Garfinkle and R.~M.~Wald,
  ``Examination of a simple example of gravitational wave memory,''
Phys.\ Rev.\ D {\bf 90}, 044060 (2014).
[arXiv:1405.6396 [gr-qc]].
}

\lref\CardosoPA{
  V.~Cardoso, O.~J.~C.~Dias and J.~P.~S.~Lemos,
  ``Gravitational radiation in D-dimensional space-times,''
Phys.\ Rev.\ D {\bf 67}, 064026 (2003).
[hep-th/0212168].
}

\lref\W{
  C.~D.~White,
  ``Factorization Properties of Soft Graviton Amplitudes,''
  JHEP {\bf 1105}, 060 (2011)
  [arXiv:1103.2981 [hep-th]].}
  
\lref\GJ{  D.~J.~Gross and R.~Jackiw,
  ``Low-Energy Theorem for Graviton Scattering,'
  Phys.\ Rev.\  {\bf 166}, 1287 (1968).}

\lref\HofmanAR{
  D.~M.~Hofman and J.~Maldacena,
  ``Conformal collider physics: Energy and charge correlations,''
JHEP {\bf 0805}, 012 (2008).
[arXiv:0803.1467 [hep-th]].
}

\lref\ZeldPoln{
Ya.~B.~Zeldovich and A.~G.~Polnarev, Sov. Astron. 18, 17 (1974).
}

\lref\BraginskyIA{
  V.~B.~Braginsky and L.~P.~Grishchuk,
  ``Kinematic Resonance and Memory Effect in Free Mass Gravitational Antennas,''
Sov.\ Phys.\ JETP {\bf 62}, 427 (1985), [Zh.\ Eksp.\ Teor.\ Fiz.\  {\bf 89}, 744 (1985)].
}

\lref\bragthorne{
V.~B.~Braginsky, K.~S.~Thorne,
``Gravitational-wave bursts with memory and experimental prospects."
Nature 327.6118 (1987): 123-125.
}

\lref\thorne{
K.~S.~Thorne,
``Gravitational-wave bursts with memory: The Christodoulou effect."
Physical Review D 45.2 (1992): 520.
}

\lref\ManasseZZ{
  F.~K.~Manasse and C.~W.~Misner,
  ``Fermi Normal Coordinates and Some Basic Concepts in Differential Geometry,''
J.\ Math.\ Phys.\  {\bf 4}, 735 (1963).
}

\lref\PoissonNH{
  E.~Poisson, A.~Pound and I.~Vega,
  ``The Motion of point particles in curved spacetime,''
Living Rev.\ Rel.\  {\bf 14}, 7 (2011).
[arXiv:1102.0529 [gr-qc]].
}

\lref\ChristodoulouCR{
  D.~Christodoulou,
  ``Nonlinear nature of gravitation and gravitational wave experiments,''
Phys.\ Rev.\ Lett.\  {\bf 67}, 1486 (1991).
}

\lref\BlanchetBR{
  L.~Blanchet and T.~Damour,
  ``Hereditary effects in gravitational radiation,''
Phys.\ Rev.\ D {\bf 46}, 4304 (1992).
}

\lref\BlanchetBR{
  L.~Blanchet and T.~Damour,
  ``Hereditary effects in gravitational radiation,''
Phys.\ Rev.\ D {\bf 46}, 4304 (1992).
}

\lref\TolishBKA{
  A.~Tolish and R.~M.~Wald,
  ``Retarded Fields of Null Particles and the Memory Effect,''
Phys.\ Rev.\ D {\bf 89}, no. 6, 064008 (2014).
[arXiv:1401.5831 [gr-qc]].
}

\lref\TolishODA{
  A.~Tolish, L.~Bieri, D.~Garfinkle and R.~M.~Wald,
  ``Examination of a simple example of gravitational wave memory,''
Phys.\ Rev.\ D {\bf 90}, 044060 (2014).
[arXiv:1405.6396 [gr-qc]].
}

\lref\AshtekarZSA{
  A.~Ashtekar,
  ``Geometry and Physics of Null Infinity,''
[arXiv:1409.1800 [gr-qc]].
}

\lref\CamanhoAPA{
  X.~O.~Camanho, J.~D.~Edelstein, J.~Maldacena and A.~Zhiboedov,
  ``Causality Constraints on Corrections to the Graviton Three-Point Coupling,''
[arXiv:1407.5597 [hep-th]].
}

\lref\BieriHQA{
  L.~Bieri and D.~Garfinkle,
  ``An electromagnetic analogue of gravitational wave memory,''
Class.\ Quant.\ Grav.\  {\bf 30}, 195009 (2013).
[arXiv:1307.5098 [gr-qc]].
}

\lref\lb{
  G.~Barnich and P.~H.~Lambert,
 ``A note on the Newman-Unti group,''
Adv.\ Math.\ Phys.\  {\bf 2012}, 197385 (2012).
[arXiv:1102.0589 [gr-qc]].
}

\lref\HeLAA{
  T.~He, V.~Lysov, P.~Mitra and A.~Strominger,
  ``BMS supertranslations and Weinberg's soft graviton theorem,''
[arXiv:1401.7026 [hep-th]].
}

\lref\FavataZU{
  M.~Favata,
  ``The gravitational-wave memory effect,''
Class.\ Quant.\ Grav.\  {\bf 27}, 084036 (2010).
[arXiv:1003.3486 [gr-qc]].
}

\lref\LudvigsenKG{
  M.~Ludvigsen,
  ``Geodesic Deviation At Null Infinity And The Physical Effects Of Very Long Wave Gravitational Radiation,''
Gen.\ Rel.\ Grav.\  {\bf 21}, 1205 (1989).
}

\lref\WinicourSKA{
  J.~Winicour,
  ``Global aspects of radiation memory,''
[arXiv:1407.0259 [gr-qc]].
}

\lref\SachsZZA{
  R.~Sachs,
  ``Asymptotic symmetries in gravitational theory,''
Phys.\ Rev.\  {\bf 128}, 2851 (1962).
}

\lref\bms{H.~Bondi, M.~G.~J.~van der Burg and A.~W.~K.~Metzner,
  ``Gravitational waves in general relativity. 7. Waves from axisymmetric isolated systems,''
Proc.\ Roy.\ Soc.\ Lond.\ A {\bf 269}, 21 (1962);
  R.~K.~Sachs,
  ``Gravitational waves in general relativity. 8. Waves in asymptotically flat space-times,''
Proc.\ Roy.\ Soc.\ Lond.\ A {\bf 270}, 103 (1962).
}

\lref\BarnichEB{
  G.~Barnich and C.~Troessaert,
  ``Aspects of the BMS/CFT correspondence,''
JHEP {\bf 1005}, 062 (2010).
[arXiv:1001.1541 [hep-th]].
}

\lref\ManasseZZ{
  F.~K.~Manasse and C.~W.~Misner,
  ``Fermi Normal Coordinates and Some Basic Concepts in Differential Geometry,''
J.\ Math.\ Phys.\  {\bf 4}, 735 (1963).
}

\lref\PoissonNH{
  E.~Poisson, A.~Pound and I.~Vega,
  ``The Motion of point particles in curved spacetime,''
Living Rev.\ Rel.\  {\bf 14}, 7 (2011).
[arXiv:1102.0529 [gr-qc]].
}

\lref\WeinbergNX{
  S.~Weinberg,
  ``Infrared photons and gravitons,''
Phys.\ Rev.\  {\bf 140}, B516 (1965).
}

\lref\KLPS{ D.~Kapec, V.~Lysov, S.~Pasterski and A.~Strominger,
  ``Semiclassical Virasoro symmetry of the quantum gravity $\cal{S}$-matrix,''
  JHEP {\bf 1408}, 058 (2014). [arXiv:1406.3312 [hep-th]].}

\lref\cs{F.~Cachazo and A.~Strominger,
  ``Evidence for a New Soft Graviton Theorem,''
  arXiv:1404.4091 [hep-th].}

\lref\LudvigsenKG{
  M.~Ludvigsen,
  ``Geodesic Deviation At Null Infinity And The Physical Effects Of Very Long Wave Gravitational Radiation,''
Gen.\ Rel.\ Grav.\  {\bf 21}, 1205 (1989).
}

\lref\MisnerQY{
  C.~W.~Misner, K.~S.~Thorne and J.~A.~Wheeler,
  ``Gravitation,''
San Francisco 1973, 1279p.
}

\lref\BieriHQA{
  L.~Bieri and D.~Garfinkle,
  ``An electromagnetic analogue of gravitational wave memory,''
Class.\ Quant.\ Grav.\  {\bf 30}, 195009 (2013).
[arXiv:1307.5098 [gr-qc]].
}

\lref\KinoshitaUR{
  T.~Kinoshita,
  ``Mass singularities of Feynman amplitudes,''
J.\ Math.\ Phys.\  {\bf 3}, 650 (1962).
}

\lref\LeeIS{
  T.~D.~Lee and M.~Nauenberg,
  ``Degenerate Systems and Mass Singularities,''
Phys.\ Rev.\  {\bf 133}, B1549 (1964).
}

\lref\WangZLS{
  J.~B.~Wang, G.~Hobbs, W.~Coles, R.~M.~Shannon, X.~J.~Zhu, D.~R.~Madison, M.~Kerr and V.~Ravi {\it et al.},
  ``Searching for gravitational wave memory bursts with the Parkes Pulsar Timing Array,''
[arXiv:1410.3323 [astro-ph.GA]].
}

\lref\HofmanAR{
  D.~M.~Hofman and J.~Maldacena,
  ``Conformal collider physics: Energy and charge correlations,''
JHEP {\bf 0805}, 012 (2008).
[arXiv:0803.1467 [hep-th]].
}

\lref\ck{
  D.~Christodoulou and S.~Klainerman,
  ``The Global nonlinear stability of the Minkowski space,''
Princeton University Press, Princeton, 1993.
}

\lref\yau{
  L.~Bieri, P.~Chen and S.~T.~Yau,
  ``The Electromagnetic Christodoulou Memory Effect and its Application to Neutron Star Binary Mergers,''
Class.\ Quant.\ Grav.\  {\bf 29}, 215003 (2012).
[arXiv:1110.0410 [astro-ph.CO]].

}

\lref\KulishUT{
  P.~P.~Kulish and L.~D.~Faddeev,
  ``Asymptotic conditions and infrared divergences in quantum electrodynamics,''
Theor.\ Math.\ Phys.\  {\bf 4}, 745 (1970), [Teor.\ Mat.\ Fiz.\  {\bf 4}, 153 (1970)]..
}

\lref\WareZJA{
  J.~Ware, R.~Saotome and R.~Akhoury,
  ``Construction of an asymptotic S matrix for perturbative quantum gravity,''
JHEP {\bf 1310}, 159 (2013).
[arXiv:1308.6285 [hep-th]].
}

\lref\WisemanSS{
  A.~G.~Wiseman and C.~M.~Will,
  ``Christodoulou's nonlinear gravitational wave memory: Evaluation in the quadrupole approximation,''
Phys.\ Rev.\ D {\bf 44}, 2945 (1991).

}
\lref\sz{  A.~Strominger and A.~Zhiboedov,
  ``Gravitational Memory, BMS Supertranslations and Soft Theorems,''
[arXiv:1411.5745 [hep-th]].

}
\lref\FavataZU{
  M.~Favata,
  ``The gravitational-wave memory effect,''
Class.\ Quant.\ Grav.\  {\bf 27}, 084036 (2010).
[arXiv:1003.3486 [gr-qc]].
}
\lref\camp{
  M.~Campiglia and A.~Laddha,
  ``New symmetries for the Gravitational S-matrix,''
[arXiv:1502.02318 [hep-th]];  ibid 

 ``Asymptotic symmetries and subleading soft graviton theorem,''
Phys.\ Rev.\ D {\bf 90}, no. 12, 124028 (2014).
[arXiv:1408.2228 [hep-th]].
}

\lref\cy{ 
  F.~Cachazo and E.~Y.~Yuan,
  ``Are Soft Theorems Renormalized?,''
  arXiv:1405.3413 [hep-th].
 }
\lref\bern{
  Z.~Bern, S.~Davies and J.~Nohle,
  ``On Loop Corrections to Subleading Soft Behavior of Gluons and Gravitons,''
  Phys.\ Rev.\ D {\bf 90}, no. 8, 085015 (2014)
  [arXiv:1405.1015 [hep-th]].
   }
  
 \lref\bailey{
  I.~Bailey and W.~Israel,
  ``Lagrangian Dynamics of Spinning Particles and Polarized Media in General Relativity,''
  Commun.\ Math.\ Phys.\  {\bf 42}, 65 (1975).}
  
%

\Title{\vbox{\baselineskip12pt}} {\vbox{
\centerline {New Gravitational Memories} }} \centerline{Sabrina Pasterski, 
Andrew Strominger and Alexander Zhiboedov} \vskip.1in \centerline{\it Center for the Fundamental Laws of Nature}\centerline{\it
Harvard University, Cambridge, MA 02138 USA}

\vskip.1in \centerline{\bf Abstract} {  The conventional gravitational memory effect is a relative displacement in the position of  two detectors induced by radiative energy flux.  We find a new type of gravitational  `spin memory'  in which beams on clockwise and counterclockwise orbits acquire a relative delay induced by radiative angular momentum flux.   It has recently been shown that the displacement memory formula is a Fourier transform in time of Weinberg's soft graviton theorem. Here we see that the spin memory formula is a Fourier transform in time of the recently-discovered subleading soft graviton theorem.   }

\Date{}

{\centerline{\bf Contents}\nobreak\medskip{\baselineskip=22pt
 \parskip=0pt\catcode`\@=11  
 \noindent {1.} {Introduction} \leaderfill{1} \par 
\noindent {2.} {Asymptotically flat metrics } \leaderfill{3} \par 
\noindent {3.} {Displacement memory effect} \leaderfill{6} \par 
\noindent {4.} {Spin memory effect} \leaderfill{6} \par 
\noindent {5.} {Spin memory and angular momentum flux} \leaderfill{8} \par 
\noindent {6.} {Equivalence to subleading soft theorem } \leaderfill{10} \par 
\noindent {7.} {An infinity of conserved charges} \leaderfill{12} \par 
\noindent Appendix {A.} {Massless particle stress-energy tensor} \leaderfill{15} \par 
\catcode`\@=12}}

 \noindent
\newsec{Introduction}

The passage of gravitational radiation  past a pair of nearby inertial detectors 
produces oscillations in their relative positions.   After the waves have passed, and spacetime locally reverts to the vacuum, the detectors in general do not return to their initial relative positions. The resulting displacement, discovered in 1974 \refs{\ZeldPoln\BraginskyIA\bragthorne\LudvigsenKG\ChristodoulouCR\WisemanSS\thorne\BlanchetBR\yau\TolishBKA-\TolishODA},  is known as the gravitational memory effect. Direct measurement of the gravitational memory may be possible in the coming years, see $e.g.$ \refs{\WangZLS , \FavataZU}.
The effect is a consequence \sz\ of the fact that the radiation induces transitions among the many BMS-degenerate \bms\ vacua in general relativity. The initial and final spacetime geometries, although both flat,  differ by a  BMS supertranslation. The displacement is proportional to the BMS-induced shift in the spacetime metric, which in turn is given by a universal formula involving moments of the asymptotic energy flux.  

   Since the initial and final metrics differ, the Fourier transform in time must have a pole at zero energy.  A universal formula for this pole was found in 1965 \WeinbergNX\  and is known as Weinberg's soft graviton theorem. The complete equivalence of the soft graviton and displacement memory formulae was demonstrated in \sz. 
   
   Recently, a new universal soft graviton formula was discovered \cs (see also \refs{\GJ,\W}) that governs not the pole but the finite piece in the expansion of soft graviton scattering about zero energy.  This was shown \KLPS\ to be a consequence of the BMS superrotations\foot{The associated symmetry group is the familiar Virasoro symmetry of euclidean two-dimensional conformal field theory \bt. This may  be usefully embedded in a larger group of all diffeomorphisms of the sphere \camp. } of \bt\ in the same sense that Weinberg's pole formula is a consequence of BMS supertranslations. 

This discovery immediately raises the question:~~is there a new kind of gravitational memory associated with superrotations and the subleading soft theorem?  In this paper we show that the answer is yes. While displacement memory is sourced by moments of the energy flux through null infinity ($\ci$), the new memory is sourced by moments of the angular momentum flux. Accordingly we call it {\it spin memory}. The spin memory effect provides a cogent operational meaning to the superrotational symmetry of gravitational scattering.  

Spin memory has a chiral structure and cannot be measured by inertial detectors. Instead, we 
consider light rays which repeatedly orbit (with the help of fiber optics or mirrors) clockwise or counterclockwise around a fixed circle in the asymptotic region. The passage of angular-momentum-carrying radiation will induce a relative time delay between the counter-orbiting light rays, resulting for example in a shift in the interference fringe.  A universal formula for this delay  is given in terms of moments of the angular momentum flux through infinity. The relative time delay for counter-orbiting light rays is the spin memory effect. It is a new kind of gravitational memory. 

This paper is organized as follows. Section 2 outlines the metric and constraint equations for asymptotically flat spacetimes.  Section 3 reviews the displacement memory effect.  Section 4 introduces the spin memory effect.  Section 5 uses the constraint equations and boundary conditions to relate this new memory effect to angular momentum flux.  Section 6 demonstrates that  the spin memory formula is the Fourier transform in time of  the subleading soft graviton formula of \cs.  Finally, in section 7, we discuss the infinite family of conserved charges associated to the infinite superrotational symmetries.  Measurements verifying the conservation laws are described.  We close with a short comment on implications for black hole information. The appendix derives several formulae concerning the angular momentum of spinning particles on null geodesics. 
\newsec{Asymptotically flat metrics }
The expansion of an asymptotically flat spacetime metric near \ip\ in retarded Bondi coordinates takes the form\foot{Our definition of $N_z$, which is proportional to the Weyl tensor (see below) is related to $N^{BT}_z$ of \bt\ by 
$4N_z= 4N^{BT}_z+C_{zz}D_zC^{zz}+{3 \over 4}\p_z(C_{zz}C^{zz}) $.}
\eqn\asymptflat{\eqalign{
d s^2 &= - d u^2 - 2 du dr + 2 r^2  \gamma_{z \bar z}dz d \bar z +2{m_B \over r} d u^2 \cr&+ 
 \bigl(r C_{zz} d z^2 +D^{z} C_{z z}dudz  +{ 1 \over r}({4\over 3}N_z-{1 \over 4}\p_z(C_{zz}C^{zz}))du dz + c.c.\bigr) +...
}}
where $u=t-r$, $\gamma_{z\zb}={2 \over (1+z\zb)^2}$ is the unit metric on $S^2$ (used to raise and lower $z$ and  $\bz$ indices), $D_z$ is the $\gamma$-covariant derivative, and subleading terms are suppressed by powers of $r$.
 The Bondi mass aspect $m_B$, the  angular momentum aspect $N_z$, and $C_{zz}$ are functions of $(u, z,\zb)$, not $r$. They are related  by the \ip\ constraint equations $G_{u u} = 8 \pi G T_{u u}^{M}$
\eqn\relaton{\eqalign{
\pa_{u} m_{B} &={1 \over 4}  \left[D_z^2 N^{z z} + D^2_{\bar z} N^{\bar z \bar z} \right] - T_{uu}, \cr T_{u u} &\equiv {1 \over 4} N_{z z} N^{z z} + 4 \pi G \lim_{r \to \infty} [r^2 T^{M}_{u u}] ,}}
 and $G_{u z} = 8 \pi G T_{u z}^{M}$
\eqn\relation{\eqalign{
\pa_{u} N_z &={1 \over 4}  \p_z \left[D_z^2 C^{z z} - D^2_{\bar z} C^{\bar z \bar z} \right] +\p_z m_B- T_{uz}, \cr
 T_{u z} &\equiv 8\pi G\lim\limits_{r\rightarrow\infty}[r^2 T^M_{uz}]-{1\over 4}D_z[C_{zz}N^{zz}]-{1\over 2}C_{zz}D_zN^{zz}, 
}}
where $N_{zz}=\p_uC_{zz}$ is the Bondi news, $T^{M}$ is the matter stress tensor and $T_{uu}$ ($T_{uz}$) is the total energy (angular momentum) flux through a given point on ${\cal I}^{+}$. The angular momentum aspect is related to the Weyl tensor component $\Psi^0_1$ on \ip\ by 
\eqn\nwy{ N_z=\lim\limits_{r\rightarrow\infty}r^3C_{zrru}.}
We also note that 
\eqn\gf{\im\Psi^0_2 =\im \lim\limits_{r\rightarrow\infty}r\gamma^{z\bz}C_{u\bz zr}=-\im [{1\over 2}D_{z}^2 C^{zz}+{1\over 4}C_{zz}N^{zz}].} 

Most of our discussion will concern \ip, but the metric expansion near $\ci^-$ in the retarded Bondi coordinate $v=t+r$ is
\eqn\asympin{\eqalign{
d s^2 &= - d v^2 + 2 dv dr + 2 r^2  \gamma_{z \bar z}dz d \bar z +2{m_B \over r} dv^2 \cr&+
 \bigl(-r C_{zz} d z^2 +D^{z} C_{z z}dvdz  +{ 1 \over r}({4\over 3}N_z+{1 \over 4}\p_z(C_{zz}C^{zz}))dv dz + c.c.\bigr) +...
}}
where here the metric perturbations are functions of $(v,z,\bz)$. The $z$ coordinate on $\ci^-$ is defined so that $(-v,z,\bz)$ is the $PT$ conjugate of $(u,z,\bz)$:~~hence they lie on the same null generator of $\ci$ and are antipodally located relative to the origin of the spacetime. The $\ci^-$ constraint equations become $G_{vv} = 8 \pi G T_{vv}^{M}$
\eqn\relatoni{\eqalign{
\pa_{v} m_{B} &=-{1 \over 4}  \left[D_z^2 N^{z z} + D^2_{\bar z} N^{\bar z \bar z} \right] + T_{vv}, \cr T_{vv} &\equiv {1 \over 4} N_{z z} N^{z z} + 4 \pi G \lim_{r \to \infty} [r^2 T^{M}_{vv}] ,}}
 and $G_{v z} = 8 \pi G T_{v z}^{M}$
\eqn\relationi{\eqalign{
\pa_{v} N_z &=-{1 \over 4}  \p_z \left[D_z^2 C^{z z} - D^2_{\bar z} C^{\bar z \bar z} \right] +\p_z m_B+T_{vz}, \cr
 T_{v z} &\equiv 8\pi G\lim\limits_{r\rightarrow\infty}[r^2 T^M_{vz}]-{1\over 4}D_z[C_{zz}N^{zz}]-{1\over 2}C_{zz}D_zN^{zz}.
}}

We use the symbol $\ci^+_-$ ($\ci^+_+$) to denote the past and future $S^2$ boundaries of \ip, 
and  $\ci^-_\pm$ for those of $\ci^-$. 
In this paper, we  consider spacetimes which decay to the vacuum 
in the far past and future $\ci^-_-$ and $\ci^+_+$. (The more general case requires an analysis of extra past and future boundary terms.) In particular,  we require 
\eqn\wdli{ N_z |_{\ci^+_+}=N_z |_{\ci^-_-}= m_B |_{\ci^+_+}=m_B|_{\ci^-_-}= 0.}
Moreover near all the boundaries $\ci^\pm_\pm$ of $\ci$, the geometry is 
in the vacuum in the sense that the radiative modes are unexcited:
\eqn\fsw{N_{zz}|_{\ci^\pm_\pm}=\im \Psi^0_2|_{\ci^\pm_\pm}=0.}
 More precisely, following Christodoulou and Klainerman \ck, we take $N_{zz}\sim |u|^{-3/2}~(|v|^{-3/2})$ on \ip\ ($\ci^-$) as well as a corresponding falloff of the stress tensor.

These conditions do  $not$ imply $C_{zz}|_{\ci^\pm_\pm}=0$.
Rather, the general solution of \fsw\ is (see e.g \StromingerJFA)
\eqn\bhj{C_{zz}=-2D_z^2C,} where $C$ is any ($u$-independent) function of $(z,\bz)$. 
These solutions are mapped to one another by BMS supertranslations and exhibit the large vacuum degeneracy in general relativity.

 As described in \StromingerJFA,~to define gravitational scattering one must specify boundary or continuity conditions on $m_B$ and $C_{zz}$ near where \ip\ and $\ci^-$ meet. The unique 
Lorentz, PT and BMS-invariant choice  is simply\eqn\cpm{ C_{zz}|_{\ci^+_-}=C_{zz}|_{\ci^-_+},~~~~~m_B|_{\ci^+_-}=m_B|_{\ci^-_+}.}
In this paper $N_z$ plays an important role and its determination from the constraint equation \relation\ also requires a continuity condition. This is a bit tricky because outside the center-of-mass frame, $N_z$ may grow linearly with $u$ or $v$ near $\ci^\pm_\pm$. It follows from \relation\ and \fsw\ that the divergent term  is exact:~~$N_z \sim u\p_z m_B$. Fortunately for us, such exact terms are irrelevant for our purposes (see section 5). We will need a continuity condition for the curl of $N_z$. \fsw, \cpm\ and the Bianchi identity  imply 
\eqn\wdl{ \p_{[z}N_{\bz]} |_{\ci^-_+}= \p_{[z}N_{\bz]} |_{\ci^+_-}.}

We are interested in the difference between the  initial and final functions $C$ in \bhj\ on $\ci$.  This can be determined by integrating the constraint \relaton\  as follows (see \sz\ for more details).
Defining
\eqn\eed{ \Delta^+ C_{zz}=C_{zz}|_{\ci^+_+}-C_{zz}|_{\ci^+_-} ,~~~~\Delta^+ m_B=m_B|_{\ci^+_+}-m_B|_{\ci^+_-}=-m_B|_{\ci^+_-},}
and using \relaton~one finds
\eqn\resultint{
D_z^2 \Delta^+ C^{z z} = 2\int d u \  T_{u u}+2 \Delta^+ m_{B} .
}
The  $\Delta^+ C$ which produces such a $\Delta^+ C_{zz}$ is obtained by
inverting $D_z^2D_{\zb}^2$:
\eqn\frfa{
\Delta^+ C(z,\zb)= \int d^2 w \gamma_{w \bar w} G(z;w)\left[ \int d u \ T_{u u} (w)  + \Delta m_{B} \right]
}
where the Green's function is given by
\eqn\manifest{\eqalign{
G(z;w) &=- {1 \over \pi}\sin^2 {\Theta \over 2}  \log\sin^2 { \Theta \over 2} , ~~~~~~~ \sin^2 {\Theta(z,w) \over 2} \equiv{|z-w|^2\over (1 + w \bar w)(1 + z \bar z) }.  
}}

An equation similar to \frfa\ may be derived for the shift of $C$ on $\ci^-$.
Adding the two equations, using the boundary condition \cpm, and defining 
\eqn\eedi{ \Delta C=\Delta^+C-\Delta^-C,}
one arrives at the simple relation 
\eqn\frf{
\Delta C(z,\zb)= \int d^2 w \gamma_{w \bar w} G(z;w)\left[ \int d u \ T_{u u} (w) - \int d v \ T_{vv} (w)   \right].
}
\newsec{Displacement memory effect}
In this section, we briefly review the standard gravitational memory effect. The passage of a finite pulse of radiation or other form of energy through a region of spacetime produces a gravitational field which moves inertial  detectors.  The final positions of a pair of nearby detectors are generically displaced relative to the initial ones according to a simple and universal formula \refs{\ZeldPoln\BraginskyIA\bragthorne\LudvigsenKG\ChristodoulouCR\WisemanSS\thorne\BlanchetBR\yau\TolishBKA-\TolishODA}, which we now review briefly.

Consider two nearby inertial detectors with proper worldline tangent vectors $t^\mu$ and relative displacement  vector $s^\mu$.  We take the worldlines to be at large $r$ and extend for infinite retarded time near \ip. $s^\mu$ evolves according to the geodesic deviation equation 
\eqn\mn{ \p_\tau^2 s^\mu=R^\mu_{~\lambda  \rho\nu}t^\lambda t^\rho s^\nu,}
where $\tau$ is the detector's proper time. At large $r$ in the geometry \asymptflat, we may approximate $\tau \sim u$, $t^\lambda\p_\lambda=\p_u$  and 
\eqn\dxl{R_{zuzu}=-{1\over 2}r \p_u^2 C_{zz}.}
It follows that 
\eqn\pkji{ \p_u^2s^\bz={\gamma^{z\bz}\over 2r }\p_u^2C_{zz}s^z. }
Integrating twice one finds, to leading order in $1 \over r$, a net change in the displacement \eqn\ssz{ \Delta^+ s^\bz= {\gamma^{z\bz}\over 2r }\Delta^+C_{zz}s^\bz,}
where $\Delta^+ C_{zz}$ is given in term of moments of the asymptotic energy flux by the second derivative of \frfa.
\ssz\ is the standard displacement memory formula. 

\newsec{Spin memory effect}
This section describes the new spin memory effect, which affects orbiting objects such as protons in the LHC, or signals exchanged by eLISA detectors. 
  Consider a circle $\CC$ of radius $L$ near \ip\ centered around a point $z_0$ on a sphere of large fixed $r=r_0$, where $L<<r_0$. This is described by 
\eqn\cdf{Z(\phi)=z_0\left[1+{{Le^{i\phi}}\over {2r_0}}{{1+z_0\bar{z}_0}\over{\sqrt{z_0 \bar{z}_0}}}\right]+\CO\left({L^2\over r_0^2}\right)}
  where $\phi \sim \phi+2\pi$. A light ray in either a clockwise or counterclockwise orbit (aided by mirrors or fiber optics) along $\CC$  starting at $\phi(0)=0$, has a  trajectory $\phi(u)$ that obeys 
\eqn\lgt{\eqalign{
ds^2&=0\cr &=1-2r_0^2\gamma_{z \bar{z}}\p_u Z\p_u\bar{Z}-2{m_{B}\over r_0}-r_0C_{zz}(\p_u Z)^2-r_0C_{\bar{z} \bar{z}}(\p_u \bar{Z})^2\cr &~~~~~~~~-[D^zC_{zz}\p_u Z+D^{\bar{z}} C_{\bar{z}\bar{z}}\p_u \bar{Z}]+...
}}
where it is here and hereafter assumed that $C_{zz}$ does not change significantly over a single period.   To this order, only the term in square brackets in~\lgt~is odd under $\p_uZ\rightarrow -\p_u Z$.  If two light rays are simultaneously set in orbit in opposite directions, the  times  at which they  return to $\phi=0$\foot{The line $\phi=0$ is a geodesic in the induced geometry of the ring \lgt\ only in the limit $r\to \infty$.  Detectors at fixed $\phi$ are `BMS detectors' of the type discussed in \sz. A finite-$r$ geodesic detector will be boosted and observe a different $\Delta P$.}  will differ by the $u$-integral of this odd term
\eqn\dsg{\Delta P=\oint_\CC \bigl( D^{z} C_{z z}dz + D^{\zb} C_{\zb \zb}d\zb\bigr).}
This formula in fact applies to any contour $\CC$, circular or not.  

At first glance, it appears from \dsg\ that the returns of the counter-orbiting light rays are desynchronized, even in the vacuum, as long as $C_{zz}$ is nonzero. In fact this is not the case. For $C_{zz}$ of the vacuum form \bhj, one readily finds that 
\eqn\dpv{\Delta P_{vacuum}=-{2 }\oint_\CC d(D^{z}D_z  C+C)=0 .}
Hence desynchronization occurs only during the passage of radiation through $\ci^+$. The total relative time delay, integrated  over all orbits is 
\eqn\ufy{\Delta^+ u={1 \over 2\pi L }\int  du \oint_\CC \bigl( D^{z} C_{z z}dz + D^{\zb} C_{\zb \zb}d\zb\bigr).}
This leads to a shift in the interference pattern between counter-orbiting light pulses. The shift is an infrared effect proportional to the $u$-zero mode of $C_{zz}$.  This is the spin memory effect. 
\newsec{Spin memory and angular momentum flux}

Displacement memory \ssz\ can be expressed as an integral of the net local asymptotic energy flux convoluted with the  Green's function \frf\ on the sphere. 
In this section we derive an analogous formulae for spin memory as a convoluted integral involving the net local asymptotic angular momentum flux. 

Taking $\p_\bz$ of the $G_{uz}$ constraint in \relation\  and $\p_z$ of the complex conjugate $G_{u\bar{z}}$ constraint gives:
\eqn\ddi{\p_z\p_{\bar{z}}m_B=\re [\p_u\p_{\bar z}N_z+\p_{\bar z}T_{uz}]}
\eqn\ddii{\im [\p_{\bar z}D_z^3C^{zz}]=2\im [\p_u\p_{\bar z}N_z+\p_{\bar z}T_{uz}].}
Multiplying \ddii\ by the Green's function
\eqn\mani{\eqalign{
{\cal G}(z;w) &=\log\sin^2 { \Theta \over 2} , ~~~ \sin^2 {\Theta(z,w) \over 2} \equiv{|z-w|^2\over (1 + w \bar w)(1 + z \bar z) }
}} which obeys \eqn\gfi{\p_z\p_\bz{\cal G}(z;w)=2\pi\delta^2(z-w)-{1\over 2}\gamma_{z\bz},}
and integrating over $d^2z$ gives:
 \eqn\ident{\pi \im [D_{w}^2C^{ww}]=-\im
\int d^2z\p_\bz{\cal G}(z;w)[\p_u N_z+T_{uz}].
}
Note that the right hand side of \ident\ is invariant under shifts $N_z\to N_z+\p_z X$  for any real $X$, so only the curl part of $N_z$ contributes.  
Integrating both sides over the disk $D_\CC$ whose boundary is $\CC$ and using Stokes' theorem, \ident\ leads to 
\eqn\contour{\pi 
\oint_{\CC}(D^{w}C_{ww}dw+D^{\bar{w}}C_{\bar{w}\bar{w}}d\bar{w})=-2\im \int_{D_{\CC}}d^2w \gamma_{w\bar{w}}\int d^2z\p_\bz{\cal G}(z;w)[\p_u N_z+T_{uz}]. 
}
Multiplying by ${ 1\over 2\pi^2 L}$ and integrating over $u$ then yields
\eqn\rrk{
\Delta^+ u=-{ 1\over \pi^2 L}\im \int_{D_{\CC}} d^2w\gamma_{w\bar{w}}\int d^2z\p_\bz{\cal G}(z;w)\left[\Delta^+ N_z+\int du T_{uz}\right].
}
where $\Delta^+ N_z\equiv N_z |_{\ci^+_+}-N_z|_{\ci^+_-}=-N_z|_{\ci^+_-}$ is the shift in the angular momentum aspect. In many applications -- for example geometries which are initially asymptotically Schwarzschild through subleading order -- this term will vanish. Moreover, if $\Delta^+N_z$ is exact, $i.e.$ 
$\Delta^+ N_z =\p_z X$ for any real $X$, no contribution to the imaginary part appears in \rrk. Hence \rrk\ depends only on the curl of 
$\Delta^+N_z$.

A similar analysis near $\ci^-$ leads to the formula
\eqn\rrkm{
\Delta^- v=-{ 1\over \pi^2 L}\im \int_{D_{\CC}} d^2w\gamma_{w\bar{w}}\int d^2z\p_\bz{\cal G}(z;w)\left[-\Delta^-N_z+\int dv T_{vz}\right].
}
where the contour $\CC$ and disk $D_\CC$ on $\ci^-$ is 
defined by the curve defined in equation \cdf\ on $\ci^-$. 
This means it will lie in the antipodal spatial direction from the origin. 
Using the continuity condition \wdl\ on $N_z$ one finds, in analogy to 
\frf, an expression relating time delays and fluxes
\eqn\xza{\Delta \tau \equiv \Delta^+u-\Delta^- v=-{ 1\over \pi^2 L}\im \int_{D_{\CC}} d^2w\gamma_{w\bar{w}}\int d^2z\p_\bz{\cal G}(z;w)\left[\int du T_{uz}
-\int dv T_{vz}\right].
}

The right hand side of \xza\ is related to  the local angular momentum flux through $\ci$. Consider the case when $T_{uz}$ arises from massless particles or localized wave packets which puncture \ip\ at points $(u_k,z_k)$.  Then, as we show in the appendix, 
\eqn\xvl{T_{uz}=
8\pi G\sum\limits_k\delta(u-u_k)\left[L_{uz}(z_k)-{i\over 2}h_k\p_z\right]{\delta^2(z-z_k)\over\gamma_{z\bar{z}}},}
together with a similar formula for $T_{vz}$.
Here $L_{uz}(z_k),$ ($h_k$) is the orbital (spin) angular momentum of the k$^{th}$ particle associated to a rotation around (boost towards) the point $z_k$ on the sphere. The leading contribution from such particles to the time delay is  
\eqn\rkii{\Delta \tau=-{8G \over \pi L}\sum\limits_k\bigl( \gamma^{z_k\bar{z}_k} \im\int_{D_\CC}d^2w\gamma_{w\bar{w}}L_{uz}(z_k)\p_{\bz_k}{\cal G}(z_k;w)+\pi h_{k\in\CC} \bigr).}
The second term in \rkii\ has the simple interpretation that an object  of spin $h_k$ passing through $\CC$ at (or near) lightspeed induces a time  
delay of order ${h_k \over L}$, with no factors of $r_0$. This can be understood as the frame-dragging effect. If $\CC$ lies a distance of order $L$ from $z_k$, the first term is typically a factor of $L \over r_0$ smaller than the second. 

Another interesting case is outgoing quadrupole radiation, with no incoming news or $T_{vz}$ on $\ci^-$.\foot{Since we are taking $m_B|_{\ci^-_-}=0$ here, the initial energy would have to enter in a spherically symmetric wave from $\ci^-$.} The displacement memory effect for configurations of this type is of potential  astrophysical interest and was analyzed in \WisemanSS. This is described by the news tensor on \ip
\eqn\tio{N_{zz}= NY^i_zY^j_z}
for some $N(u)$.  As an example, take $i=j$, $Y_i^z=z$ and
\eqn\czz{N={\alpha \over (2\pi)^{1/4}}\p_u e^{-u^2+i\omega u}.}
The resulting angular momentum flux  obeys \eqn\ing{\int du T_{uz}={3i\over 2}\alpha^2\omega \p_z{z^2\bar{z}^2\over (1+z\bar{z})^4},}
so that
\eqn\inm{\im \int du \int d^2z\p_\bz{\cal G}(z;w)T_{uz}=3\pi\alpha^2\omega \left[{1\over 30}-{w^2\bar{w}^2\over (1+w\bar{w})^4}\right],}
using \gfi.
The quadrupole contribution to the time delay around a contour $\CC$ becomes
\eqn\ghj{ \Delta^+ u={3\alpha^2 \omega \over \pi L} \int_{D_{\CC}} d^2w\gamma_{w\bar{w}} \left[{w^2\bar{w}^2\over (1+w\bar{w})^4}-{1\over30}\right].}
For typical choices of $\CC$ such that the area of $D_\CC$ is order $L^2/r_0^2$, we have $\Delta^+ u \sim {\alpha^2 L \over r_0^2}$.

\newsec{Equivalence to subleading soft theorem }

In the sixties, Weinberg \WeinbergNX\  showed that scattering amplitudes in any theory with gravity exhibit universal poles as the energy $\omega$ of any external graviton is taken to zero. 
Recently \refs{\cs,\GJ,\W,\KLPS,\camp,\cy,\bern} it has been shown  that  the finite, subleading term in the $\omega \to 0$ expansion also exhibits universal behavior.  The coefficient of the  $leading$ pole was shown in \sz\ to be related by a  timeline Fourier transform of the expression for displacement memory.
We now show that the $subleading$ term is a Fourier transform of  the expression for spin memory. 

The subleading  soft graviton theorem  is a universal  relation between ($n+1$)-particle (with one soft graviton) and $n$-particle  tree-level quantum scattering amplitudes \refs{\cs}\ 
\eqn\qet{
{1\over 2}(\lim_{\omega \to 0+}+\lim_{\omega \to 0-})\CA_{n+1}\bigl(p_1,...p_n;  (\omega q,\epsilon_{\mu\nu})\bigr)=S^{(1)}_{\mu\nu} \eps^{\mu\nu} \CA_{n}\bigl(p_1,...p_n\bigr),}
where
\eqn\rft{
S^{(1)}_{\mu\nu}={i\kappa\over 2}\sum_{k=1}^n{p_{k(\mu}J_{k\nu)\lambda}q^\lambda \over q\cdot p_k} 
}
and $\kappa=\sqrt{32\pi G}$.  
 In this expression, the parentheses denote $(\mu,\nu)$ symmetrization, $q=(\omega,\omega \hat  q)$ with $\hat q^2=1$ is the four-momentum, and $\epsilon_{\mu\nu}$ is the  transverse-traceless polarization tensor of the  graviton. We define incoming particles to have negative $p^0$ and take $\omega$ positive for an outgoing graviton.  $\mu,\nu$ indices refer to asymptotically Minkowskian coordinates given in terms of retarded coordinates as \eqn\ddp{\eqalign{
x^0&=u+r,  \cr
  x^1+ix^2&={2rz \over 1+z\bz},\cr
  x^3&= {r(1-z\bz) \over 1+z\bz}, }}  
  or in terms of advance coordinates as 
 \eqn\ddpi{\eqalign{
x^0&=v-r,  \cr
  x^1+ix^2&=-{2rz \over 1+z\bz},\cr
  x^3&= -{r(1-z\bz) \over 1+z\bz}. }}  
The symmetrized limit in \qet\  projects out the leading Weinberg pole, leaving the subleading finite term of interest here. The linearized  expectation value of the asymptotic metric fluctuation produced in the  $n$-particle  scattering process obeys the semiclassical  momentum space formula
\eqn\frrr{\eqalign{(
\lim_{\omega \to 0+}+\lim_{\omega \to 0-})h_{\alpha\beta }(\omega,q)&= \epsilon_{\alpha\beta}(\lim_{\omega \to 0+}+\lim_{\omega \to 0-}){  \CA_{n+1}\bigl(p_1,...p_n;  (\omega q ,\epsilon_{\mu\nu})\bigr)\over\CA_{n}\bigl(p_1,...p_n\bigr)} \cr 
&=  ~{i\kappa} \epsilon_{\alpha\beta}\epsilon^{\mu \nu} \sum_{k=1}^n{p_{k\mu} J_{k\nu\lambda} q^\lambda \over q\cdot p_k}.}}
In the last line, and in similar expressions below,  an expectation value of the expression involving the differential operator $J_{k\nu\lambda}$ acting in the matrix element in $\CA_n$ is implicit.

Expression \frrr\  characterizes  linearized fields by their momenta  whereas the new memory formula \xza\ is given in terms of
$\ci$ values of fields. These are simply related. 
Using the large-$r$ stationary phase approximation  as in \refs{\KLPS,\HeLAA}
\eqn\fourier{
\int du C_{zz}(u,\hat q)-\int dv C_{zz}(v,\hat q) = -(
\lim_{\omega \to 0+}+\lim_{\omega \to 0-}){i \kappa\over 8\pi }\p_zX^\mu\p_z X^\nu h_{\mu \nu} (\omega, \hat q ) ,}
where  $X^\mu\equiv \lim\limits_{r\rightarrow\infty}{x^\mu\over r}$ and the unit vector $\hat q$ is viewed as a coordinate on $\ci$.  The hard particle momenta  $p_k$, the soft graviton momentum $q$, and complex polarization $\e^{-\mu\nu}=\e^{-\mu}\e^{-\nu}$ are given in terms of the  points $z_k$ and $z$ at which they  arrive on the 
on the asymptotic $S^2$ and their energies $E_k,\omega$ 
\eqn\iok{\eqalign{p_k^\mu&={E_k\over1+z_k\bar{z}_k}\left(1+z_k\bar{z}_k,\bz_k+z_k,i(\bar{z}_k-z_k),1-z_k\bar{z}_k\right),\cr
q^\mu&={\omega \over 1+z\bar{z}}\left(1+z\bar{z},\bar{z}+z,i(\bar{z}-z),1-z\bar{z}\right),\cr
\epsilon^{-\mu}&={1\over\sqrt{2}}(z,1,i,-z).
}} 
Rewriting the soft formula \frrr\ in terms of the variables in \fourier\ and \iok, 
defining $\hat{S}^{(1)}_{zz}\equiv\p_zX^\mu\p_z X^\nu S^{(1)}_{\mu\nu}$, and
acting with $D_z^2$ gives \KLPS\
\eqn\tot{\eqalign{\im\big[ \int du D_z^2 C_{\bz\bz}- \int dvD_z^2 C_{\bz\bz}\big]&={\kappa\over 8\pi}[D_{\bz}^2 \hat{S}^{(1)}_{zz}-D_z^2 \hat{S}^{(1)}_{\bz\bz}].\cr}}
The formulae for the angular momentum and stress energy of a particle emerging at $z_k$ in the appendix enables this to be rewritten: 
\eqn\exp{\eqalign{\im\big[\int du D_z^2 C^{zz}-&\int dv D_z^2 C^{zz}\big]\cr
&=-8G\sum\limits_k \gamma^{z_k\bar{z}_k}\im \left[L_{uz}(z_k)\p_{\bz_k}{\cal G}(z_k;z)+ {i\over 2}h_k\p_{z_k}\p_{\bz_k}{\cal G}(z_k;z)\right]\cr& =-{1\over \pi}\im
\int d^2w\p_{\bar{w}}{\cal G}(w;z)\left[\int duT_{uw}-\int dv T_{vw}\right].
}}
where total angular momentum conservation has been used.  To summarize, the subleading soft graviton theorem \cs, after some differentiation, change of notation and Fourier transform, becomes the formula for the contour integrals characterizing spin memory. 
 \newsec{An  infinity of conserved charges}

A conserved charge enables one to determine the outcome of a measurement on \ip\ from a measurement on $\ci^-$. For example, for a process which begins and ends in a vacuum,  
the total integrated outgoing energy flux across \ip\ equals the incoming energy flux across $\ci^-$. 
In this section we describe an $infinite$ set of \ip\ measurements -- one for every contour $\CC$ -- whose outcome is determined by a measurement on $\ci^-$. 

Conserved charges on \ip\ may be obtained from any moment of the curl $\p_{[\bz} N_{z]}$ on $\ci^+_-$. Consider for example  \eqn\uk{Q(z,\bz)=i\p_{[\bz} N_{z]}(z,\bz)|_{\ci^+_-}.}
This can be written using the constraints as an integral over a null generator of  \ip. Integrating by parts then gives the $\ci^+$ expression
\eqn\usk{Q(z,\bz)=-i\int du\left[{1 \over 4}  \p_\bz \p_z \left[D_z^2 C^{z z} - D^2_{\bar z} C^{\bar z \bar z} \right] - \p_{[\bz}T_{z]u}\right].}
On the other hand using the continuity condition \wdl\ and integrating by parts, one finds the $\ci^-$ expression
\eqn\uskk{Q(z,\bz)=-i\int dv\left[{1 \over 4}  \p_\bz \p_z \left[D_z^2 C^{z z} - D^2_{\bar z} C^{\bar z \bar z} \right] - \p_{[\bz}T_{z]v}\right].}Equating \usk\ and \uskk\ gives the conservation law:
\eqn\cvn{\eqalign{&\int du\left[{1 \over 4}  \p_\bz \p_z \left[D_z^2 C^{z z} - D^2_{\bar z} C^{\bar z \bar z} \right] - \p_{[\bz}T_{z]u}\right]\cr
&~~~~~~~~~~~~~~~~~=\int dv\left[{1 \over 4}  \p_\bz \p_z \left[D_z^2 C^{z z} - D^2_{\bar z} C^{\bar z \bar z} \right] - \p_{[\bz}T_{z]v}\right].\cr}}
This conservation law equates the stress energy flux through and a zero mode of the metric fluctuations along a null generator of \ip\ to the same quantities on the $PT$-conjugate generator of $\ci^-$. There is one such law for 
every null generator. 

The meaning of the conservation law \cvn\ is a bit obscured by the fact that  zero modes of metric fluctuations are hard to measure. However, in the preceding we have found that the time delay effectively measures a particular combination of the zero modes. Our main formula \xza\ can be rephrased as a conservation laws for the charge \eqn\tok{Q_\CC \equiv-{1\over \pi^2 L}\im \int_{D_{\CC}} d^2w\gamma_{w\bar{w}}\int d^2z\p_\bz{\cal G}(z;w)N_z |_{\ci^+_-},}
which, as noted above, involves only the curl of $N_z$. Using \wdl\  $Q_\CC$ may be rewritten as 
 a $\ci^-$ charge 
\eqn\tk{Q_\CC\equiv-{1\over \pi^2 L}\im \int_{D_{\CC}} d^2w\gamma_{w\bar{w}}\int d^2z\p_\bz{\cal G}(z;w)N_z |_{\ci^-_+}.} Integrating by parts, using the constraints to express  
\tok\ and \tk\ as integrals over $\ci^\pm$, and equating the two expressions yields
\eqn\rok{\eqalign{&\Delta^+ u+{1\over \pi^2 L}\im \int_{D_{\CC}} d^2w\gamma_{w\bar{w}}\int d^2z\p_\bz{\cal G}(z;w)\int du T_{uz}\cr
&~~~=\Delta^- v+{1\over \pi^2 L}\im \int_{D_{\CC}} d^2w\gamma_{w\bar{w}}\int d^2z\p_\bz{\cal G}(z;w)\int dv T_{vz}.}}

Thus if we measure the component $T_{vz}$ of the radiative stress-energy flux and the time delay on $\CC$ at past null infinity, we can determine a moment of $T_{uz}$ and the time delay of an antipodally-located contour at future null infinity. There are infinitely many such conservation laws -- one for every contour $\CC$ -- which infinitely constrain the scattering process. 

Should it persist to the quantum theory, this infinity of conservation laws has 
considerable implications for the black hole information puzzle. The output of the black hole evaporation process, as originally computed by Hawking, is constrained only by energy-momentum, angular momentum and charge conservation.  Imposing the infinity of conservation laws \rok\ (together with a second infinity arising from BMS invariance \StromingerJFA) will greatly constrain the outgoing Hawking radiation.  These constraints follow solely from low-energy symmetry considerations, and do not invoke any microphysics. It would be interesting to understand how the semiclassical computation of black hole evaporation must be modified to remain  consistent with these symmetries. 

~

\centerline{\bf Acknowledgements}
We are grateful to D. Christodoulou, S. Gralla, T. He, D. Kapec and P. Mitra for useful conversations. This work was supported in part by NSF grant 1205550. 
\vfil\break
 
 \appendix{A}{Massless particle stress-energy tensor}
 
We start with the trajectory of a massless point particle:
\eqn\xmu{x^\mu(\tau)={p^\mu\over E}\tau+b^\mu}
where $p^\mu$ is the particle's four momentum, with $p_\mu p^\mu=0$.  $b^\mu=(0,b^i)=x^\mu(0)$ describes the impact parameter of the straight-line trajectory relative to the spacetime origin.  The orbital angular momentum of this trajectory is:
\eqn\lmn{L^{\mu\nu}=b^\mu p^\nu-p^\mu b^\nu,}
which implies 
\eqn\ffe{b^\mu={1\over E}L^{\mu 0}.} The total angular momentum is 
\eqn\jmn{
J^{\mu\nu}=L^{\mu\nu}+S^{\mu\nu}
}
where $S_{\mu\nu}$ is the intrinsic spin.   
The large $\tau$ behavior of the trajectory \xmu\ is:
\eqn\tra{\eqalign{r(\tau)&=\tau+{1\over E^2}p^\mu L_{u\mu}+\CO(\tau^{-1})\cr
u(\tau)&=-{1\over E^2}p^\mu L_{u\mu}+\CO(\tau^{-1})\cr
z(\tau)&={p^1+ip^2\over E+p^3}+{1\over E}L_{u}^{~z} +\CO(\tau^{-2})\cr
}}
where we have used $L_{0\mu}=L_{u\mu}$ and ${1\over E}L_{u}^{~z}$ is $\CO(\tau^{-1})$.
 The matter stress-energy tensor of the point particle is \bailey:
\eqn\tmn{T^M_{\mu\nu}(y^\rho)=E\int d\tau \dot{x}_\mu\dot{x}_\nu{\delta^4(y^\rho-x^\rho(\tau))\over \sqrt{-g}}-\nabla^\rho\int d\tau S_{\rho(\mu} \dot{x}_{\nu)}{\delta^4(y^\rho-x^\rho(\tau))\over \sqrt{-g}}.}
Using $S_{z\bar{z}}=ir^2\gamma_{z\bar{z}}h$ near a particle with helicity $h$, a collection of point particles obeys:
\eqn\tco{\eqalign{\lim\limits_{r\rightarrow\infty} r^2T^M_{uu}&=\sum\limits_kE_k\delta(u-u_k){\delta^2(z-z_k)\over\gamma_{z\bar{z}}}\cr
 \lim\limits_{r\rightarrow\infty} r^2T^M_{uz}&=\sum\limits_k\delta(u-u_k)\left[L_{uz}(z_k)-{i\over 2}h_k\p_z\right]{\delta^2(z-z_k)\over\gamma_{z\bar{z}}}
 }}
where $u_k\equiv -{1\over E_k^2}p_k^\mu L_{ku\mu}$ and 
\eqn\lz{\eqalign{L_{uz}(z_k)&\equiv \lim\limits_{r\rightarrow\infty}{1\over r}{\p x_k^\mu\over \p {u_k}}{\p x_k^\nu\over\p {z_k}} L_{k\mu\nu}\cr
&={b_k^1(1-\bar{z}_k^2)-ib_k^2(1+\bar{z}_k^2)-2b_k^3\bar{z}_k\over (1+z_k\bar{z}_k)^2}E_k.}}

\listrefs

\bye